\pdfoutput=1
\documentclass[11pt,twoside]{article}
\usepackage{graphicx,epsfig,natbib,epstopdf}
\usepackage{CS18}

\begin{document}

\title{HST FUV monitoring of TW Hya}
\author{
H. M. G\"unther$^1$, 
N. S. Brickhouse$^1$, A. K. Dupree$^1$, S. J. Wolk$^1$, P. C. Schneider$^2$ , G. J. M. Luna$^3$}

\affil{$^1$Harvard-Smithsonian Center for Astrophsics, 60 Garden St, Cambridge, MA 02138,USA}
\affil{$^2$Hamburger Sternwarte, Gojenbergsweg 112, 21029 Hamburg, Germany}
\affil{$^3$IAFE/Conicet, Casilla de Correo 67, Buenos Aires, Argentina}
\begin{abstract}
Classical T Tauri stars (CTTS) are young ($< 10$ Myr), cool stars that actively
accrete matter from a disk. They show strong, broad and asymmetric, atomic FUV emission lines. Neither the width, nor the line profile is understood. Likely, different mechanisms influence the line profile; the best candidates are accretion, winds and stellar activity. We monitored the C~{\sc iv} 1548/1550 \AA{} doublet in the nearby, bright CTTS TW Hya with the Hubble Space Telescope Cosmic Origin Spectrograph (HST/COS) to correlate it with i) the cool wind, as seen in COS NUV Mg II line profiles, ii) the photometric period from joint ground-based monitoring, iii) the accretion rate as determined from the UV continuum, and iv) the H$\alpha$ line profile from independent ground-based observations. The observations span 10 orbits distributed over a few weeks to cover the typical time scales of stellar rotation, accretion and winds. Here we describe a model with intrinsically asymmetric C IV lines.

\end{abstract}

\section{Introduction}
Young cool stars are surrounded by a circumstellar
disk. As long as they actively accrete from the disk, they are called
classical T~Tauri star (CTTS). For a general review about CTTS see \citet{2013AN....334...67G} in
the proceeding of the previous Cool Stars workshop. Here we present new data on the closest
CTTS known to date, the star TW~Hya \citep[57~pc,][]{1998MNRAS.301L..39W}. In CTTS, the disk does not reach down to the
stellar surface but it is truncated at a few stellar radii. From that point on,
material is funneled along the lines of the stellar magnetic fields. TW~Hya is
one of the oldest known CTTS, and thus the
accretion rate from the disk onto the star is low. Still, the mass is
accelerated almost to free-fall velocities (about 500~km~s$^{-1}$) and when it
hits the stellar surface a strong accretion shock forms. This shock is the
energy source for a range of phenomena that distinguish a CTTS spectrum from a
main-sequence star.

\section{The accretion shock in CTTS}
The temperature of the accretion shock is
high enough to cause X-ray emission. \citet{2002ApJ...567..434K} noticed unusual line ratios in
the He-like oxygen triplet of O~{\sc vii}, that indicate the region where the
X-rays are formed is much denser than typical stellar
coronae. \citet{acc_model} performed numerical simulations and matched a model
that mixed an accretion shock with a stellar corona to the observed X-ray
spectrum. Yet, \citet{2010ApJ...710.1835B} could show that the densities observed in
Ne~{\sc ix} are higher than in O~{\sc vii}, contrary to the expectation of
shock models. This led them to propose that only the Ne~{\sc ix} emission is
really formed in the post-shock region that extends from the shock front down
into the stellar atmosphere, while the O~{\sc vii} emission comes from material
that is heated in the shock front but then escapes ``sideways'' into the
transition region and lower corona. Because material in the transition region
experiences lower pressures than material deeper in the stellar atmosphere,
its density will be lower. 

Here, we will show how the same scenario might also explain the line
profiles observed in atomic FUV emission lines. From the FUV to the IR, the
accretion shock also powers an additional continuum, the so-called ``veiling''
that reduces the relative strength of the absorption lines. According to the model
of \citet{calvetgullbring} this component originates in regions close to the
cooling zone that are heated by the energetic radiation of the hot,
post-shock layers. This component is
often fitted as a black-body of a few thousand K, but recently
\citet{2012AstL...38..649D,2013AstL...39..389D} have argued that it
fills up stellar absorption lines, too.

We describe that the accretion shock is one viable model for the FUV emission
lines, but we do not claim that it is the only one. In the full publication
that we are currently preparing, we will also examine a hot wind model as suggested in \citet{2005ApJ...625L.131D} and \citet{2014ApJ...789...27D}.

\section{Observations}
We monitored the C~{\sc iv} 1548/1550~\AA{} doublet in the TW Hya with the
Hubble Space Telescope Cosmic Origin Spectrograph (HST/COS) for 10 orbits
distributed over a few weeks to cover the typical time scales of stellar
rotation, accretion and winds.
The ObsID is 12315 and all automatic data processing worked normally.  Data were retrieved from the archive and processed using
custom python scripts.
We measured the radial velocities of molecular hydrogen lines and interstellar
absorption lines and found a velocity precision better than 6~km~s$^{-1}$ for
the FUV channel and better than 15~km~s$^{-1}$ for the NUV channel. For the
purposes of this contribution, this is sufficient; we will correct the velocity scale using this
measurement in our paper (in prep.).

\section{Results}
\begin{figure}[ht!]
\centering
\includegraphics[width=90mm]{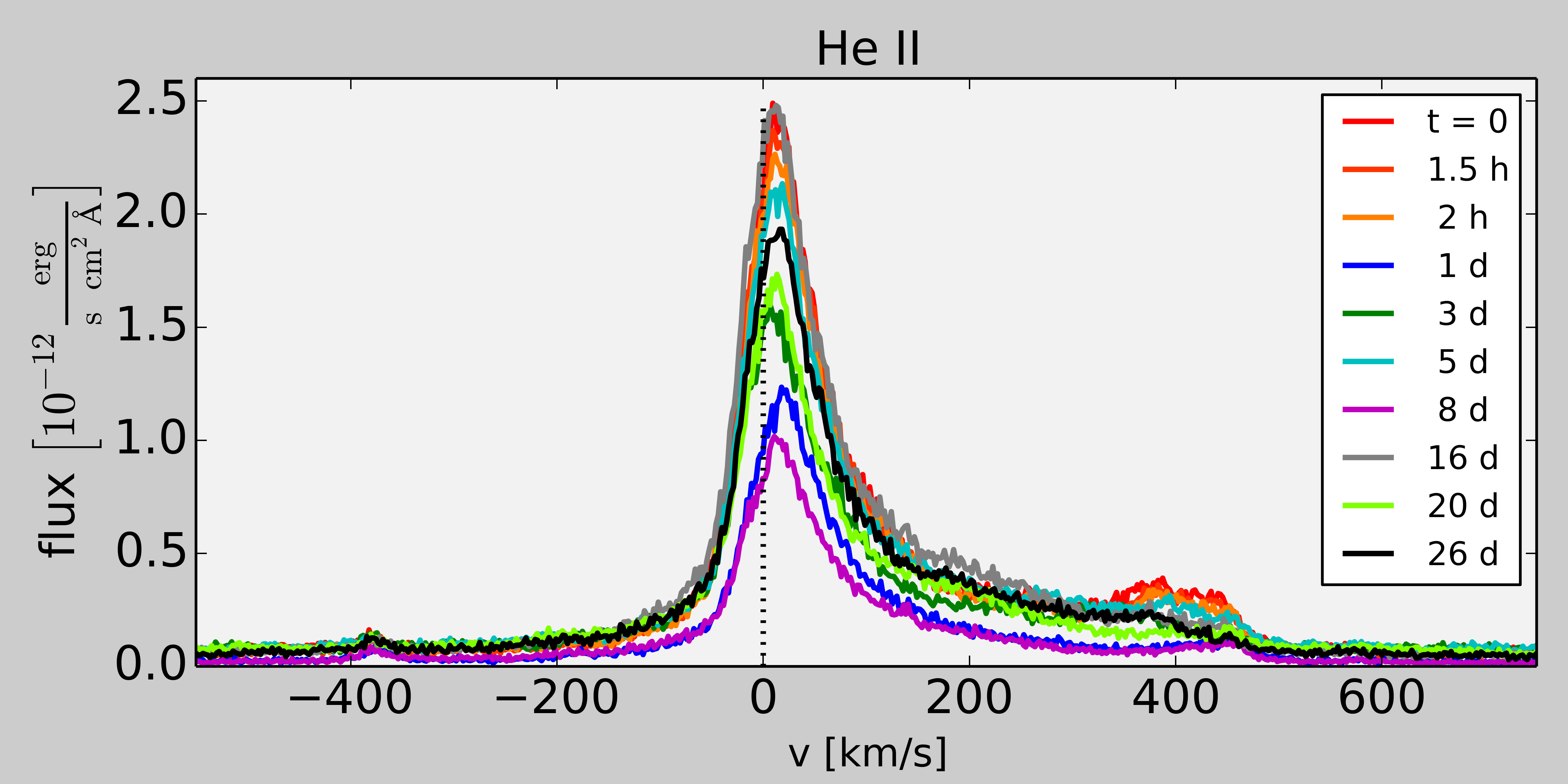}
\caption{Line profiles of the He~{\sc ii} 1640~\AA{} line. Each color belongs to one observation. The legend in the figure shows how long after the first each observation was taken.}
\label{fig:HeII}
\end{figure}

\begin{figure}[ht!]
\centering
\includegraphics[width=90mm]{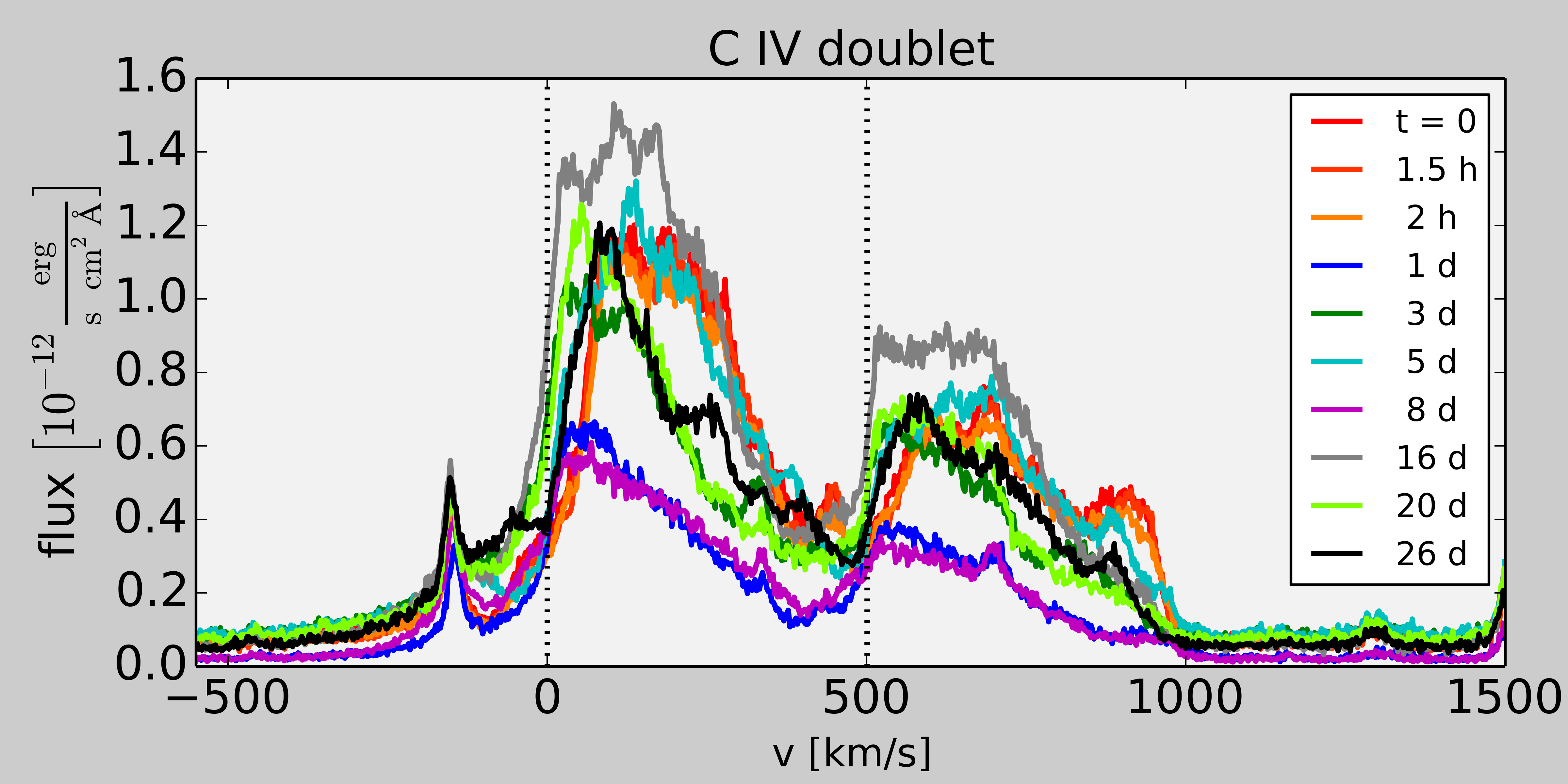}
\caption{Line profiles of the C~{\sc iv} doublet at 1548/1550~\AA{}. The dotted lines in the figure indicate the rest wavelengths for both lines of the doublet.}
\label{fig:CIV}
\end{figure}

\begin{figure}[ht!]
\centering
\includegraphics[width=90mm]{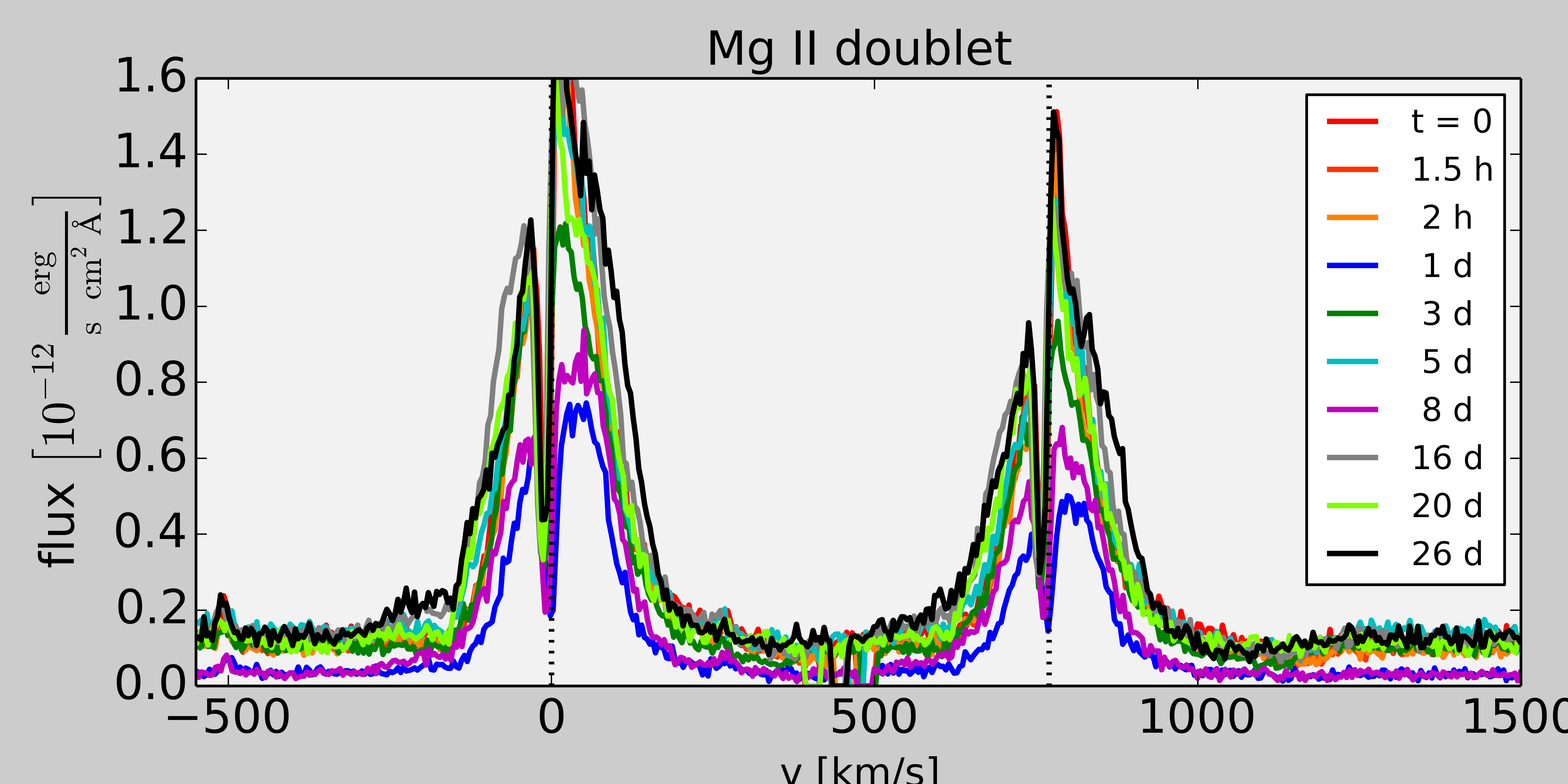}
\caption{Line profiles of the Mg~{\sc ii} 2796/2804~\AA{} doublet.}
\label{fig:MgII}
\end{figure}

Figure~\ref{fig:HeII} to \ref{fig:MgII} show the line profiles of the strongest FUV emission lines. In addition to these atomic lines, there are a large number of ro-vibrational molecular hydrogen lines. These lines are formed in the inner disk through photo-excitation by strong FUV emission lines  \citep{2002ApJ...572..310H,2004ApJ...607..369H}. Their variability will be analyzed in a forthcoming publication. 

\begin{figure}[ht!]
\centering
\includegraphics[width=90mm]{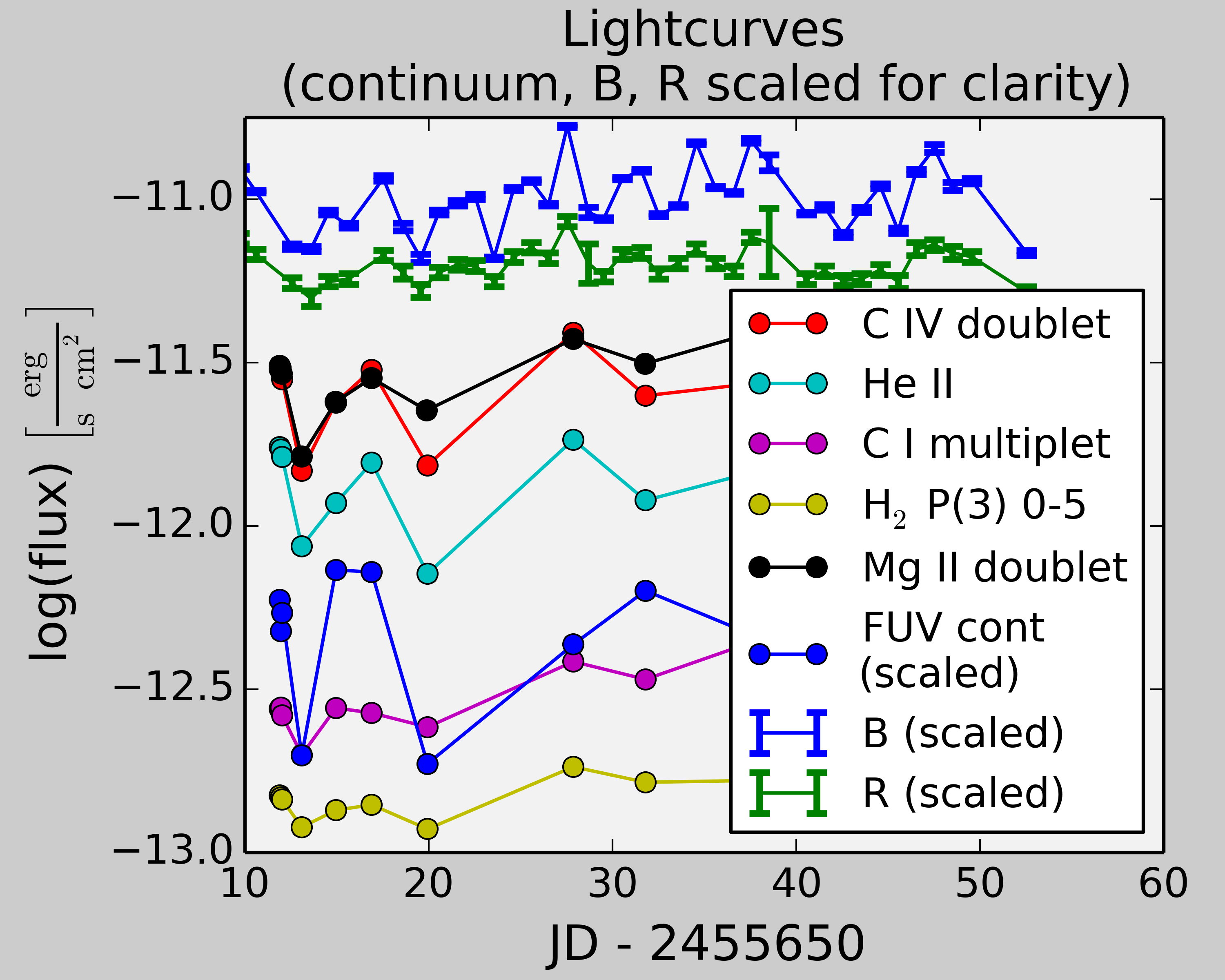}
\caption{Lightcurves in the FUV emission lines, the continuum and from ground-based monitoring.}
\label{fig:lc}
\end{figure}

In all three figures there is clearly a continuum. We have performed a number of tests (e.g.\ we have extracted only the time in the earth shadow and compared that to the day-time data) and found no hint for an instrumental origin, we thus conclude that the continuum comes from TW~Hya. It varies in phase with the emission lines. When the emission lines are stronger, so is the continuum (Fig.~\ref{fig:lc}); the amplitude of the continuum variation is about a factor of three.

Figure~\ref{fig:HeII} shows the He~{\sc ii} line profile, which is the simplest of the strong FUV lines. It peaks close to the rest wavelength and its peak position is virtually constant in all observations as is the shape of the line. The only exception is an emission bump around 400~km/s that appears in the first three spectra.

A similar bump is seen in the C~{\sc iv} line in figure~\ref{fig:CIV} for the same spectra at the same velocity. C~{\sc iv} is a doublet, where the blue line should be twice as strong as the red line if there is no optical depth. In the figure, the velocity scale is calculated with respect to the blue line. There is little flux on the blue side of the line. The flux then rises sharply close to the rest wavelength (the exact value varies between 0 and 100 km/s). The shape of the blue side of the line is highly variable, but in all cases the line extends to about 500~km/s. 

The Mg~{\sc ii} doublet in figure~\ref{fig:MgII} has a simple shape again. It is almost symmetric, centered on the rest wavelength and has a very narrow interstellar absorption line superposed.

In the following we concentrate on the interpretation of the C~{\sc iv} line.

\section{Interpretation}
If we want to interpret the C~{\sc iv} line in the context of an accretion shock emission model, then we expect a line that has no blue wing (since all material moves away from us into the star and the far side of the star is hidden from view), has a first peak close to the rest wavelength and a second peak close to the pre-shock velocity \citep{2003ARep...47..540L}.
We can make three important observations from the line profile:
\begin{enumerate}
\item The line rises always at slightly red-shifted velocities between 0 and +100 km/s. We do not have enough points in time to find a
pattern. We conclude that absorption happens between the accretion spot and us.
\item The $H_2$ emission line at -100 km/s is not absorbed \citep{2007ApJ...655..345J}, thus there can be no hot wind which contains C~{\sc iv} between the disk and us. A full discussion of a hot wind model is beyond the scope of this contribution, but hot winds might still be possible, if they have the right temperature profile.
\item We see no absorption below the continuum (e.g.\ a P-Cyg profile), thus there is no hot outflow with C~{\sc iv} between the continuum emission (presumably the accretion shock) and us.
\end{enumerate}

\section{Model}
We think that a model with five components can explain the observed profiles. At first sight that might seems like a very complicated model, but all of the five components are already well established from observations in other wavelength bands. 
\begin{description}
\item[Transition region] \citet{2013ApJS..207....1A} observed that even non-accreting T Tauri stars show some C~{\sc iv} emission, presumably from active regions and the transition region. However, this component is negligibly weak for TW~Hya.
\item[Pre-shock accretion column] Gas just before the impact on the stellar surface moves with the free-fall velocity (about 400 km/s for TW Hya) and is already ionized due to the radiation field in this region \citep{2003ARep...47..540L}.
\item[Post-shock cooling zone] Most of the emission comes from this region. After the shock front the material slows down, but continues to move deeper into the star. Thus, the peak of the emission is redshifted by a small velocity only \citep{2003ARep...47..540L}. In this region, the flow becomes turbulent and the line width exceeds 200~km/s.
\item[Mass spill-up] Some post-shock gas can escape from the accretion funnel
  into the transition region according to simulations
  \citep{2010A&A...510A..71O}. This mass is clearly hot enough to contain
  C~{\sc iv} and will cause line absorption. Its velocity is uncertain, but
  must be lower than the free-fall velocity. If this flow is time variable (because the material escapes from a turbulent region), then the absorption can switch on and off, depending on the amount of material in the line-of-sight and thus the point where the observed flux rises can change between 0 and 100 km/s.
\item[Heated photosphere] Last, the heated photosphere around the accretion shock will emit as a blackbody as discussed above. This can explain the continuum. Since this is a large emission region, it is not affected by the absorption due to the previous component.
\end{description}

\section{Summary}
TW~Hya is an accreting CTTS. We present the first results of our analysis of ten COS spectra, taken in the FUV and NUV. Atomic lines are strong and highly variable in the observation. We show that accretion shock emission is one possible model for the observed line variability.

\acknowledgments{This work is funded by NASA/HST grant NASA-HST-GO-12315.01.
Based on observations with the Hubble Space Telescope and SMARTS. The Analysis uses IPython \citep{IPython} and astropy \citep{2013A&A...558A..33A}.
}


\normalsize
\begin{thebibliography}{18}
\expandafter\ifx\csname natexlab\endcsname\relax\def\natexlab#1{#1}\fi

\bibitem[{{Ardila} {et~al.}(2013){Ardila}, {Herczeg}, {Gregory}, {Ingleby},
  {France}, {Brown}, {Edwards}, {Johns-Krull}, {Linsky}, {Yang}, {Valenti},
  {Abgrall}, {Alexander}, {Bergin}, {Bethell}, {Brown}, {Calvet}, {Espaillat},
  {Hillenbrand}, {Hussain}, {Roueff}, {Schindhelm}, \&
  {Walter}}]{2013ApJS..207....1A}
{Ardila}, D.~R., {et~al.} 2013, \apjs, 207, 1

\bibitem[{{Astropy Collaboration} {et~al.}(2013){Astropy Collaboration},
  {Robitaille}, {Tollerud}, {Greenfield}, {Droettboom}, {Bray}, {Aldcroft},
  {Davis}, {Ginsburg}, {Price-Whelan}, {Kerzendorf}, {Conley}, {Crighton},
  {Barbary}, {Muna}, {Ferguson}, {Grollier}, {Parikh}, {Nair}, {Unther},
  {Deil}, {Woillez}, {Conseil}, {Kramer}, {Turner}, {Singer}, {Fox}, {Weaver},
  {Zabalza}, {Edwards}, {Azalee Bostroem}, {Burke}, {Casey}, {Crawford},
  {Dencheva}, {Ely}, {Jenness}, {Labrie}, {Lim}, {Pierfederici}, {Pontzen},
  {Ptak}, {Refsdal}, {Servillat}, \& {Streicher}}]{2013A&A...558A..33A}
{Astropy Collaboration} {et~al.} 2013, \aap, 558, A33

\bibitem[{{Brickhouse} {et~al.}(2010){Brickhouse}, {Cranmer}, {Dupree}, {Luna},
  \& {Wolk}}]{2010ApJ...710.1835B}
{Brickhouse}, N.~S., {Cranmer}, S.~R., {Dupree}, A.~K., {Luna}, G.~J.~M., \&
  {Wolk}, S. 2010, \apj, 710, 1835

\bibitem[{{Calvet} \& {Gullbring}(1998)}]{calvetgullbring}
{Calvet}, N., \& {Gullbring}, E. 1998, \apj, 509, 802

\bibitem[{{Dodin} \& {Lamzin}(2012)}]{2012AstL...38..649D}
{Dodin}, A.~V., \& {Lamzin}, S.~A. 2012, Astronomy Letters, 38, 649

\bibitem[{{Dodin} \& {Lamzin}(2013)}]{2013AstL...39..389D}
---. 2013, Astronomy Letters, 39, 389

\bibitem[{{Dupree} {et~al.}(2014){Dupree}, {Brickhouse}, {Cranmer}, {Berlind},
  {Strader}, \& {Smith}}]{2014ApJ...789...27D}
{Dupree}, A.~K., {Brickhouse}, N.~S., {Cranmer}, S.~R., {Berlind}, P.,
  {Strader}, J., \& {Smith}, G.~H. 2014, \apj, 789, 27

\bibitem[{{Dupree} {et~al.}(2005){Dupree}, {Brickhouse}, {Smith}, \&
  {Strader}}]{2005ApJ...625L.131D}
{Dupree}, A.~K., {Brickhouse}, N.~S., {Smith}, G.~H., \& {Strader}, J. 2005,
  \apjl, 625, L131

\bibitem[{{G{\"u}nther}(2013)}]{2013AN....334...67G}
{G{\"u}nther}, H.~M. 2013, Astronomische Nachrichten, 334, 67

\bibitem[{{G{\"u}nther} {et~al.}(2007){G{\"u}nther}, {Schmitt}, {Robrade}, \&
  {Liefke}}]{acc_model}
{G{\"u}nther}, H.~M., {Schmitt}, J.~H.~M.~M., {Robrade}, J., \& {Liefke}, C.
  2007, \aap, 466, 1111

\bibitem[{{Herczeg} {et~al.}(2002){Herczeg}, {Linsky}, {Valenti},
  {Johns-Krull}, \& {Wood}}]{2002ApJ...572..310H}
{Herczeg}, G.~J., {Linsky}, J.~L., {Valenti}, J.~A., {Johns-Krull}, C.~M., \&
  {Wood}, B.~E. 2002, \apj, 572, 310

\bibitem[{{Herczeg} {et~al.}(2004){Herczeg}, {Wood}, {Linsky}, {Valenti}, \&
  {Johns-Krull}}]{2004ApJ...607..369H}
{Herczeg}, G.~J., {Wood}, B.~E., {Linsky}, J.~L., {Valenti}, J.~A., \&
  {Johns-Krull}, C.~M. 2004, \apj, 607, 369

\bibitem[{{Johns-Krull} \& {Herczeg}(2007)}]{2007ApJ...655..345J}
{Johns-Krull}, C.~M., \& {Herczeg}, G.~J. 2007, \apj, 655, 345

\bibitem[{{Kastner} {et~al.}(2002){Kastner}, {Huenemoerder}, {Schulz},
  {Canizares}, \& {Weintraub}}]{2002ApJ...567..434K}
{Kastner}, J.~H., {Huenemoerder}, D.~P., {Schulz}, N.~S., {Canizares}, C.~R.,
  \& {Weintraub}, D.~A. 2002, \apj, 567, 434

\bibitem[{{Lamzin}(2003)}]{2003ARep...47..540L}
{Lamzin}, S.~A. 2003, Astronomy Reports, 47, 540

\bibitem[{{Orlando} {et~al.}(2010){Orlando}, {Sacco}, {Argiroffi}, {Reale},
  {Peres}, \& {Maggio}}]{2010A&A...510A..71O}
{Orlando}, S., {Sacco}, G.~G., {Argiroffi}, C., {Reale}, F., {Peres}, G., \&
  {Maggio}, A. 2010, \aap, 510, A71+

\bibitem[{P\'erez \& Granger(2007)}]{IPython}
P\'erez, F., \& Granger, B.~E. 2007, Computing in Science and Engineering, 9,
  21

\bibitem[{{Wichmann} {et~al.}(1998){Wichmann}, {Bastian}, {Krautter},
  {Jankovics}, \& {Rucinski}}]{1998MNRAS.301L..39W}
{Wichmann}, R., {Bastian}, U., {Krautter}, J., {Jankovics}, I., \& {Rucinski},
  S.~M. 1998, \mnras, 301, L39+

\end{thebibliography}
\end{document}